\documentclass[prd,aps,twocolumn,amsmath,amssymb,nofootinbib,preprintnumbers]
{revtex4}

\usepackage{graphicx}
\usepackage{dcolumn}
\usepackage{bm}
\usepackage{amsmath}
\usepackage{amsfonts}
\usepackage{ifthen}
\usepackage{psfrag}

\def\ls{\mathrel{\lower4pt\vbox{\lineskip=0pt\baselineskip=0pt
           \hbox{$<$}\hbox{$\sim$}}}}
\def\gs{\mathrel{\lower4pt\vbox{\lineskip=0pt\baselineskip=0pt
           \hbox{$>$}\hbox{$\sim$}}}}
\def\drawbox#1#2{\hrule height#2pt

\hbox{\vrule width#2pt height#1pt \kern#1pt
              \vrule width#2pt}
              \hrule height#2pt}

\def\Asym#1#2{\vcenter{\vbox{\drawbox{#1}{#2}
              \kern-#2pt       
              \drawbox{#1}{#2}}}}

\def\beq{\begin{equation}
\def\eeq{\end{equation}}}

\newcommand{\be}{\begin{equation}}
\newcommand{\ee}{\end{equation}}
\newcommand{\bea}{\begin{eqnarray}}
\newcommand{\eea}{\end{eqnarray}}

\begin{document}

\title{MSSM inflation, dark matter, and the LHC}

\author{Rouzbeh Allahverdi$^{1}$}
\author{Bhaskar Dutta$^{2}$}
\author{Yudi Santoso$^{3}$}

\affiliation{$^{1}$~Department of Physics and Astronomy, University of New Mexico, Albuquerque, NM 87131, USA \\
$^{2}$~Department of Physics and Astronomy, Texas A\&M University, College Station, TX 77843-4242, USA\\
$^{3}$~Department of Physics and Astronomy, University of Kansas, Lawrence, KS 66045-7582, USA}

\begin{abstract}
Inflation can occur near a point of inflection in the potential of flat directions of the Minimal Supersymmetric Standard Model. In this paper we elaborate on the complementarity between the bounds from Cosmic Microwave Background measurements, dark matter and particle physics phenomenology in determining the underlying parameters of MSSM inflation by specializing to the Minimal Supergravity scenario. We show that the future measurements from the Large Hadron Collider in tandem with all these constraints will significantly restrict the allowed parameter space. We also suggest a new perspective on the fine tuning issue of MSSM inflation. With quantum corrections taken into account, the necessary condition between the soft supersymmetry breaking parameters in the inflaton potential can be satisfied at scales of interest without a fine tuning of their boundary values at a high scale. The requirement that this happens at the inflection point determines a dimensionless coupling, which is associated with a non-renormalizable interaction term in the Lagrangian and has no bearing for phenomenology, to very high accuracy.
\end{abstract}
MIFPA-10-15 \\ April, 2010
\maketitle


\section{Introduction}

Inflation is the dominant paradigm of the early universe cosmology to solve the problems of the hot big-bang model and create the seeds for structure formation. Although observations strongly support a period of superluminal expansion~\cite{WMAP}, a successful realization of inflation within particle physics has remained as a challenge. Recently it has been shown~\cite{AEGM,AEGJM,AKM} that inflation can happen within the Minimal Supersymmetric Standard Model (MSSM) and its minimal extensions.

In these models inflation occurs near a point of inflection along a $D$-flat direction~\cite{GKM} in the scalar potential of Supersymmetric (SUSY) partners of quarks and leptons (called squarks and sleptons respectively). The scale of inflation is very low, $H_{\rm inf} \sim {\cal O}(100~{\rm MeV})$, and the requirement to generate density perturbations of the correct size singles out two $D$-flat directions, which consist of squarks and sleptons respectively, as the inflaton candidates. Since the inflaton belongs to the
observable sector~\footnote{Low scale inflection point inflation can also happen in the hidden sector~\cite{ADS}.}, its couplings to matter and its decay products are known, therefore it is possible
to track the thermal history of the universe right from the end of inflation. Also, inflation is compatible with SUSY dark matter~\cite{ADM1} (and even a unified origin of inflation and dark matter may emerge~\cite{ADM2})~\footnote{For a review on MSSM inflation, see~\cite{Rouzbeh}.}.

MSSM inflation has remarkable features. The mere fact that the inflaton is related to squarks and sleptons implies that it can be tested outside cosmology. 
This is quite interesting because it gives the first example of an inflationary model with predictions for phenomenology, and hence experiments other than measurements from the Cosmic Microwave Backgrond (CMB) are needed to identify the allowed parameter space of inflation. Another feature, which is due to the fact that inflation occurs near a point of inflection, is that, unlike other models of inflation, MSSM inflation can give rise to a wide range of the scalar spectral index~\cite{LK,AEGJM} including the whole range allowed by the WMAP data~\cite{WMAP}. This, coming as a virtue, also raises an issue. The robustness comes at the expense of a finely tuned relationship between two dimensionful parameters (i.e. the soft SUSY breaking mass and the $A$-term of the flat direction that plays the role of the inflaton). The seriousness of the issue is that this fine tuning is not protected by a symmetry and needs to be performed to several orders in perturbations theory.

In this work we investigate these two issues in more detail. We point out that once quantum corrections are taken into account, the necessary condition between the the soft mass and $A$-term can be satisfied without a fine tuning in their input values at a high scale like the Grand Unified Theory (GUT) scale. One actually needs to tune a dimensionless coupling that controls the Vacuum Expectation Value (VEV) of the inflection point, to ensure that the relationship between the soft SUSY breaking parameters is satisfied at the right scale. This coupling represents a non-renormalizable interaction term that has no bearing for phenomenology, and its only role is to lead to successful inflation in MSSM.

We also demonstrate the complementarity of cosmological and phenomenological bounds in restricting the parameter space of MSSM inflation by performing a detailed study for the Minimal Supergravity (mSUGRA) scenario. We show that bounds from SUSY dark matter and mass measurements at the Large Hadron Collider (LHC), as well as those from the muon anomalous magnetic moment and rare decays, significantly restrict the allowed region of the parameter space. More data from different experiments can therefore pin down the model parameters in the future.

The organization of this paper is as follows. In Section II we give a brief recount of inflection point inflation in MSSM. In Section III we discuss the parameter space of MSSM inflation and constraints from the cosmological density perturbations. We discuss the fine tuning issue in light of radiative corrections in Section IV, and suggest that it can be considered as tuning of a dimensionless parameter that is relevant only for inflation. We then specialize to the mSUGRA scenario in Section V and show how various bounds (dark matter, sparticle mass spectrum, muon anomalous magnetic moment, etc) significantly restrict the allowed parameter space. We close the paper by concluding remarks in Section VI.

\section{Inflection point inflation in MSSM}

We start with a brief recount of inflation in MSSM~\cite{AEGM,AEGJM}. The inflaton candidates are $udd$ and $LLe$ flat directions defined by
\beq \label{inflaton}
{\hat \phi} = {{\tilde u} + {\tilde d} + {\tilde d} \over \sqrt{3}} ~ ~ ~ , ~ ~ ~ {\hat \phi} = {{\tilde L} + {\tilde L} + {\tilde e} \over \sqrt{3}}.
\eeq
Here ${\tilde u},~{\tilde d}$ are the right-handed (RH) up- and down-type squarks respectively, and ${\tilde L},~{\tilde e}$ are the left-handed (LH) sleptons and RH charged sleptons respectively. These flat directions are lifted by non-renormalizable superpotential terms of order $6$ that can be parameterized as $\lambda {\hat \phi}^6/M^3_{\rm P}$~\cite{GKM}, where $\lambda$ is a dimensionless coupling and $M_{\rm P}$ is the reduced Planck mass.

After writing ${\hat \phi} = \phi~{\rm exp}(i \theta)/\sqrt{2}$, and minimizing the potential along the angular direction $\theta$,
the scalar potential is found to be~\cite{AEGM,AEGJM}
\beq \label{scpot}
V(\phi) = {1\over2} m^2_\phi\, \phi^2 - A {\lambda\phi^6 \over 6\,M^{6}_{\rm P}} + \lambda^2
{{\phi}^{10} \over M^{6}_{\rm P}}\,,
\eeq
where $m_\phi$ and $A$ are the soft breaking mass and the $A$-term respectively ($A$ is a positive quantity after its phase is absorbed by a redefinition of $\theta$).

If
\beq \label{dev}
\alpha^2 \equiv {1 \over 4} \left({A^2 \over 40 m^2_{\phi}} - 1 \right) \ll 1 \, ,
\eeq
there exists a point of inflection
\begin{eqnarray}
&&\phi_0 = \left({m_\phi M^{3}_{\rm P}\over \lambda \sqrt{10}}\right)^{1/4} \, , \label{infvev} \\
\, \nonumber
\end{eqnarray}
in $V(\phi)$ at which
\begin{eqnarray}
\label{pot}
&&V(\phi_0) = \frac{4}{15}m_{\phi}^2\phi_0^2  \, , \\
\label{1st}
&&V^{\prime}(\phi_0) = 4 \alpha^2 m^2_{\phi} \phi_0  \, , \\
\label{2nd}
&&V^{\prime \prime}(\phi_0) = 0 \, , \\
\label{3rd}
&&V^{\prime \prime \prime}(\phi_0) = 32\frac{m_{\phi}^2}{\phi_0} \, ,
\end{eqnarray}
to the leading order in $\alpha^2$~\footnote{To be precise, the values of $m_\phi,~A,~\lambda$ in these expressions are at the scale $\phi_0$. For a detailed discussion, see~\cite{AEGJM}.}.
We note that for weak scale SUSY the inflection point has a sub-Planckian VEV.

The potential is extremely flat in the vicinity of the inflection point and the slow-roll parameters $\epsilon \equiv (M^2_{\rm P}/2)(V^{\prime}/V)^2$ and $\eta \equiv M^2_{\rm P}(V^{\prime \prime}/V)$ are smaller than $1$ within the interval
$\vert \phi - \phi_0 \vert \sim (\phi^3_0 / 60 M^2_{\rm P})$.
In consequence, inflation will occur if $\phi$ is sufficiently close to $\phi_0$ and has a negligible kinetic energy~\footnote{For the initial condition of MSSM inflation, see~\cite{Initial}.}. The Hubble expansion rate during inflation is given by
$H_{\rm inf} \simeq (m_{\phi}\phi_0/45 M_{\rm P})$.
The amplitude of density perturbations $\delta_H$ and the scalar spectral index $n_s$ are given by~\cite{AEGJM,LK}:
\beq \label{ampl}
\delta_H = {8 \over \sqrt{5} \pi} {m_{\phi} M_{\rm P} \over \phi^2_0}{1 \over \Delta^2}
~ {\rm sin}^2 [{\cal N}_{\rm COBE}\sqrt{\Delta^2}]\,, \eeq
and
\beq \label{tilt}
n_s = 1 - 4 \sqrt{\Delta^2} ~ {\rm cot} [{\cal N}_{\rm COBE}\sqrt{\Delta^2}],
\eeq
where
\beq \label{Delta}
\Delta^2 \equiv 900 \alpha^2 {\cal N}^{-2}_{\rm COBE} \Big({M_{\rm P} \over \phi_0}\Big)^4\,.
\eeq
${\cal N}_{\rm COBE}$ is the number of e-foldings between the time when observationally relevant perturbations are generated and the end of inflation. In MSSM inflation the universe enters a radiation-dominated phase immediately after the end of inflation~\cite{AEGJM}, which results in~\cite{LL}
\beq \label{NCOBE}
{\cal N}_{\rm COBE} \simeq 66.9 + {1 \over 4} {\rm ln}\left({V(\phi_0) \over M^4_{\rm P}}\right).
\eeq

\section{Parameter space of MSSM inflation}

It is seen from Eq.~(\ref{scpot}) that the inflaton potential has three parameters: $m_\phi,~\lambda$ and $A$, denoting the soft mass of the flat direction that plays the role of the inflaton, coupling of the non-renormalizable superpotential term that lifts the flat direction, and the $A$-term associated with the higher order superpotential. The soft mass and $A$-term should satisfy the relation in Eq.~(\ref{dev}) in order to have an inflection point in $V(\phi)$, thus a sufficiently flat potential, that is suitable for inflation at sub-Planckian field values. For weak scale SUSY, which we consider here, $m_\phi,~A \sim 100~{\rm GeV}-1$ TeV. However, the coupling $\lambda$ has no bearing for phenomenology since it governs a non-renormalizable interaction term. Its order of magnitude depends on the underlying high scale physics that induces the superpotential term $\lambda {\hat \phi}^6/M^3_{\rm P}$. One can in general have the following cases:

\begin{itemize}
\item \underline{Case 1:} {This term is compatible with all gauge symmetries and is induced by physics above the Planck scale (quantum gravity or string theory). In this case we may expect $\lambda \sim {\cal O}(1)$.}
\item \underline{Case 2:} {This term is induced by physics above the Planck scale, but is not compatible with all gauge symmetries. It originates from a superpotential term of higher order, $\lambda^{\prime} \phi^6 \chi^n/M^{6+n}_{\rm P}$ with $\lambda^{\prime} \sim {\cal O}(1)$. Then, spontaneous symmetry breaking at a scale $M$ (for example, GUT scale $M_{\rm GUT}$) results in $\langle \chi \rangle \sim M$, which yields the term $\lambda {\hat \phi}^6/M^3_{\rm P}$ where $\lambda = \lambda^{\prime} (M/M_{\rm P})^n$. In this case we can expect $\lambda \ll 1$~\footnote{A small $\lambda$ can also be obtained in the intersecting $D$-brane models where the magnitude of the coupling is determined from the area obtained from intersecting points.}.}
\item \underline{Case 3:} {This term is induced by new physics at a scale $M$ much lower than the Planck scale (for example, GUT scale $M_{\rm GUT}$). The term in its original form looks like $\lambda^{\prime} {\hat \phi}^6/M^3$ with $\lambda^{\prime} \sim {\cal O}(1)$. Once it is cast into the form $\lambda {\hat \phi}^6/M^3_{\rm P}$, we find $\lambda = \lambda^{\prime} (M_{\rm P}/M_{\rm GUT})^3$. In this case we can expect $\lambda \gg 1$.}
\end{itemize}

Since $\phi_0$ has a mild dependence on $\lambda$, i.e. $\phi_0 \propto \lambda^{-1/4}$, see Eq.~(\ref{infvev}), all of these cases with very different values of $\lambda$ can be compatible with MSSM inflation. Therefore we treat $\lambda$ as a free parameter with arbitrary order of magnitude. As we will see, it can eventually be determined from a combination of various experiments.
This will then shed light on the underlying physics that induces the higher order term whose only role is to give rise to successful inflation within MSSM.

We note that the Lagrangian parameters $A,~\lambda$ can be traded for $\phi_0,~\Delta^2$ through Eqs.~(\ref{dev},\ref{infvev},\ref{pot},\ref{Delta},\ref{NCOBE}). In fact, it is more appropriate to use $m_\phi,~\phi_0$ and $\Delta^2$ as parameters since they appear in the expressions for the two inflation  observables $\delta_H$ and $n_s$~(\ref{ampl},\ref{tilt}). The scalar spectral index $n_s$ is mainly sensitive to $\Delta^2$ and has only a mild logarithmic dependence on $m_\phi$ and $\phi_0$ through ${\cal N}_{\rm COBE}$. On the other hand, the amplitude of density perturbations $\delta_H$ mainly depends on $m_\phi$ and $\phi_0$. Thus one can use the value of $\delta_H$ and the $2 \sigma$ range for $n_s$ from WMAP 7-year data~\cite{WMAP} to find the allowed region in the $m_\phi-\phi_0$ plane and the allowed range for $\Delta^2$ respectively.

\begin{figure}
\begin{center}
\includegraphics[width=.48\textwidth]{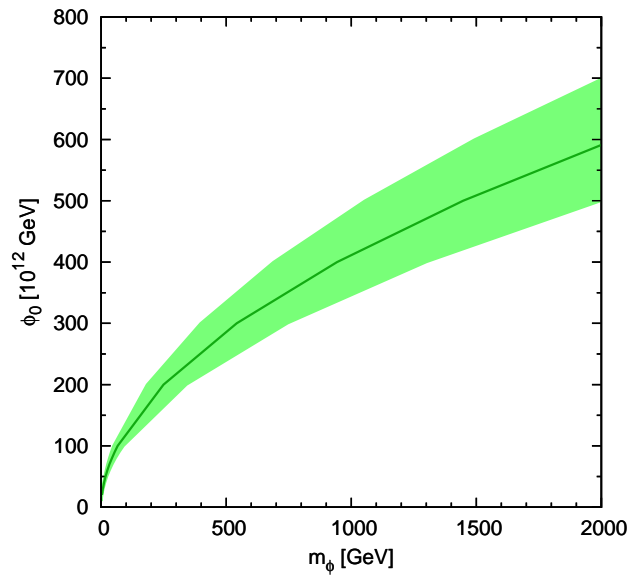}
\end{center}
\vskip -0.1in
\caption{The green band shows the acceptable region in the $m_\phi-\phi_0$ plane where MSSM inflation generates density perturbations compatible with the $2 \sigma$ region allowed by WMAP data. The dark curve at the center corresponds the central value of $n_s$.}
\label{nsdH}
\end{figure}

We show the acceptable region of $m_\phi-\phi_0$ plane that is compatible with successful inflation in Fig.~\ref{nsdH}. The ranges for $m_\phi$ and $\phi_0$ shown in the figure correspond to experimentally interesting cases where sparticle masses are accessible at the LHC~\footnote{Note that $m_\phi$ depends the soft masses of squarks (sleptons) in the case of $udd$ ($LLe$) flat direction as the inflaton.}:
\bea
{\cal O} (100 ~ {\rm GeV}) \ls & m_\phi & \ls {\cal O}(1~{\rm TeV}) \, , \nonumber \\
10^{14} ~ {\rm GeV} \ls & \phi_0 & \ls 10^{15} ~ {\rm GeV} \, . \label{mphiphi0}
\eea
Within this allowed region we must have
\be \label{Delta2}
\Delta^2 \sim {\cal O}(10^{-6}),
\ee
in order to have an acceptable $n_s$.

It is important to note that measuring two observables from inflation cannot lead to determination of all the three parameters. This is evident from Fig.~\ref{nsdH} where observational limits on $\delta_H$ and $n_s$ are translated into a band that extends through the $m_\phi-\phi_0$ plane. Therefore other measurements are needed to pinpoint an acceptable point in the parameter space. This is the main topic of our discussion in this paper.

\section{The fine tuning issue and a new perspective}

Inflection point inflation is very robust in that it can generate $n_s$ within a broad range, while keeping $\delta_H$ unchanged, by a slight change in the model parameters. For weak scale SUSY, this occurs by having $\Delta^2 \sim {\cal O}(10^{-6})$, which implies from Eq.~(\ref{dev}) that $\alpha \sim 10^{-10}$. This amounts to a severe fine tuning in the ratio of $A$ and $m_\phi$. The more serious problem is that such a fine tuning, if made at the tree level, is not stable under radiative corrections. The reason being that the existence of an inflection point in the potential is seemingly unrelated to any symmetry.

However radiative corrections can turn into a virtue here. At the tree level, the ratio $A^2/40 m^2_\phi$ is a constant that does not depend on the flat direction VEV. If it satisfies Eq.~(\ref{dev}) with $\alpha \sim 10^{-10}$, then there will be a point of inflection in the potential that is suitable for a successful inflation. Otherwise the ensuing inflation will not be compatible with observations or, if $\alpha$ is too large, there will be no inflation at all.

But it is important to note that because of quantum corrections, $m_{\phi}$ and $A$ depend on the flat direction VEV, which sets the mass of particles in the relevant quantum loops. Once we know the boundary values of $m_\phi$ and $A$, usually given at the GUT scale, we can find their values at any other scale by using the relevant renormalization group equations (RGEs). These equations (at one loop) read
\begin{eqnarray}
\mu{dm_{\phi}^2\over{d\mu}}&=&{-1\over{6\pi^2}}({4}{M_3^2}g_3^2+ {2 \over {5}}{M_1^2}g_1^2)\, , \nonumber \\
\mu{dA\over{d\mu}}&=&{-1\over{4\pi^2}}({16\over 3}{M_3}g_3^2+ {8\over {5}}{M_1}g_1^2)\, , \label{udd}
\end{eqnarray}
for the $udd$ flat direction, and
\begin{eqnarray}
\mu{dm_{\phi}^2\over{d\mu}}&=&{-1\over{6\pi^2}}({3 \over 2}{M_2^2}g_2^2+{9\over {10}}{M_1^2}g_1^2)\, , \nonumber \\
\mu{dA\over{d\mu}}&=&{-1\over{4\pi^2}}({3 \over 2}{M_2}g_2^2+{9\over {5}}{M_1}g_1^2)\, , \label{LLe}
\end{eqnarray}
for the $LLe$ flat direction. Here $M_1,~M_{2},~M_3$ and $g_1,~g_2,~g_3$ are the $U(1)_Y,~SU(2)_W,~SU(3)_C$ gaugino masses, and gauge couplings respectively. Note that $m_\phi$ is related to the soft masses of squarks or sleptons according to $m^2_\phi = (m^2_{\tilde u} + m^2_{\tilde u} + m^2_{\tilde d})/3$ and $m^2_\phi = (m^2_{\tilde L} + m^2_{\tilde L} + m^2_{\tilde e})/3$, in the two cases respectively.

The running of $m_\phi$ and $A$ implies that $\alpha$ is also a scale-dependent quantity. As shown in Eq.~(\ref{mphiphi0}), the phenomenologically interesting range of $\phi_0$ is $10^{14}-10^{15}$ GeV, which is below $M_{\rm GUT}$. Therefore we need the condition $\alpha \sim 10^{-10}$ to be satisfied at some scale $\mu$ within this range. This can happen, as a result of running, even if $\alpha \gg 10^{-10}$ at $M_{\rm GUT}$.

\begin{figure}
\begin{center}
\includegraphics[width=.48\textwidth]{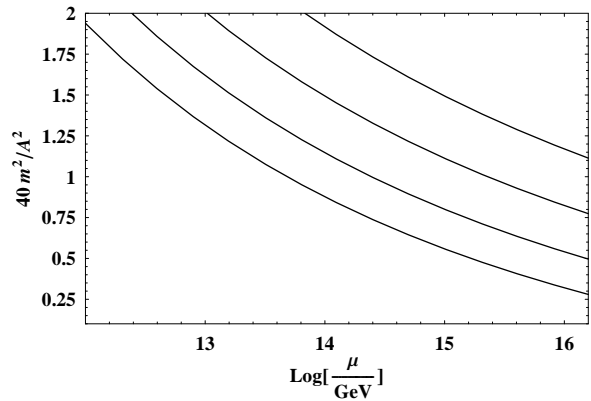}
\end{center}
\vskip -0.1in
\caption{The ratio $(40 m^2_\phi/{A^2})$ as a function of ${\rm Log}[{\mu \over{\rm GeV}}]$ in the case of $udd$ flat direction. The curves are for $M_{\rm GUT}$ boundary values $m_\phi$= 150, 200, 250, 300 GeV (respectively from left to right), and $A=1.6$ TeV.}
\label{udd1}
\end{figure}

\begin{figure}
\begin{center}
\includegraphics[width=.48\textwidth]{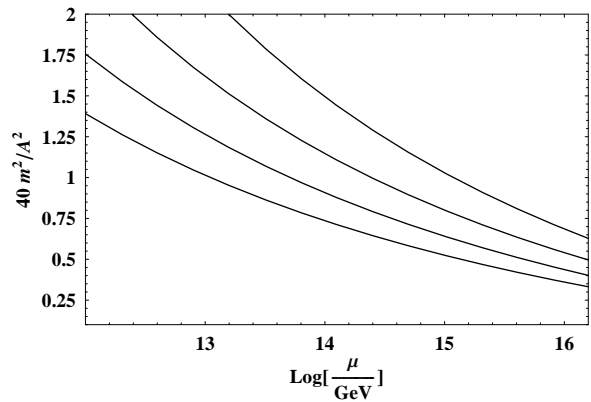}
\end{center}
\vskip -0.1in
\caption{The ratio $({40 m^2_\phi}/{A^2})$ as a function of ${\rm Log}[{\mu\over{\rm GeV}}]$ in the case of $udd$ flat direction. The curves are for $M_{\rm GUT}$ boundary values $A_{udd}$=1.6, 1.8, 2.0, 2.2 TeV (respectively from top to bottom), and $m_{\phi}=400$ GeV.}
\label{udd2}
\end{figure}

This is clearly demonstrated in Figs.~\ref{udd1}, \ref{udd2} where we show the value of $(40 m^2_\phi/A^2)$ as a function of scale $\mu$ in the case of $udd$ flat direction. In Fig.~\ref{udd1}, we plot $(40 m^2_{\phi}/A^2) $ vs ${\rm Log}[\mu/GeV]$ for various $m_{\phi}$ in the range of 150 to 300 GeV and fixed $A=1.6$~TeV. This range is allowed by low energy phenomenology and will easily be accessible in the initial run of the LHC. In Fig.~\ref{udd2}, we fix $m_\phi$ to be 400 GeV but vary A from 1.6 to 2.2 TeV. We find that in both cases $A^2 = 40 m^2_\phi$ is achieved within the range $\mu = 10^{14}-10^{15}$ GeV for $\alpha$ as large as ${\cal O}(1)$ at the GUT scale~\footnote{A similar situation happens for the $LLe$ flat direction, but the acceptable range of $\alpha$ at $M_{\rm GUT}$ is smaller because of the slower running of $m_\phi$ and $A$ in this case, which is due to the absence of gluino loops in this case.}. The situation is summarized in a dotted plot in Fig.~4, where the scale at which $A^2 = 40 m^2_\phi$ is shown vs the boundary value of $40 m^2_\phi/A^2$.

Therefore the condition $\alpha \sim 10^{-10}$ can be dynamically satisfied at $\mu = 10^{14}-10^{15}$ GeV without a severe fine tuning between the boundary values of $A$ and $m_\phi$. However, in order to have a successful inflation, this scale must coincide with the VEV of the inflection point in the potential $\phi_0$, given in Eq.~(\ref{infvev}), to an accuracy of ${\cal O}(\alpha^2)$. Once we know the values of $m_\phi$ and $A$ at $M_{\rm GUT}$, the only unknown parameter in~(\ref{infvev}) is the dimensionless coupling $\lambda$, which can be determined from the coincidence of the two scales.

This leads to a new perspective on the fine tuning issue of MSSM inflation:\\
\\
{\it The actual issue can be considered as that of tuning on the dimensionless coupling $\lambda$, not tuning two dimensionful parameters $m_\phi$ and $A$ against each other}.\\
\\
Thus, starting at $M_{\rm GUT}$, no severe tuning between the boundary values of $A$ and $m_\phi$ is needed. After using the RGEs in Eqs.~(\ref{udd},\ref{LLe}), we can find the scale at which $(A^2/40 m^2_\phi) - 1 \ls O(10^{-10})$. This scale must coincide with the inflection point VEV $\phi_0$, from which we can find $\lambda$ from Eq.~(\ref{infvev}). This is a one loop calculation, which can be extended to higher order loops. In order to ensure the precise coincidence of the scales, $\lambda$ must be determined to an accuracy of ${\cal O}(\alpha^2)$, where $\alpha \sim 10^{-10}$ is needed for successful inflation. Therefore this procedure must be repeated up to several loops.

We emphasize that the condition to have an inflection point suitable for inflation does not impose a fine tuned relation between $A$ and $m_\phi$ a priori. It rather requires that $\lambda$ be determined to a very high accuracy. This is conceivable since $\lambda$ has no bearing for phenomenology, and its primary role is to lead to a successful inflation.

One final comment is in order. In principle, it is possible that the soft SUSY breaking parameters are introduced at a scale below $M_{\rm GUT}$~\footnote{Predictions from this type of scenario have been studied in~\cite{oliveellissandick}.}.
If this scale coincides with the inflection point VEV $\phi_0$, then the boundary values of $A$ and $m_\phi$ must be tuned against each other to very high accuracy $\alpha \sim 10^{-10}$ in order for inflation to happen. Note that the fine tuning is a tree level issue in this case. The ratio $A^2/40 m^2_\phi$ is controlled by a single parameter in the SUSY breaking sector (for example, see~\cite{EMN}), which is determined very accurately from the requirement of having a suitable inflection point.

\begin{figure}
\begin{center}
\includegraphics[width=.48\textwidth]{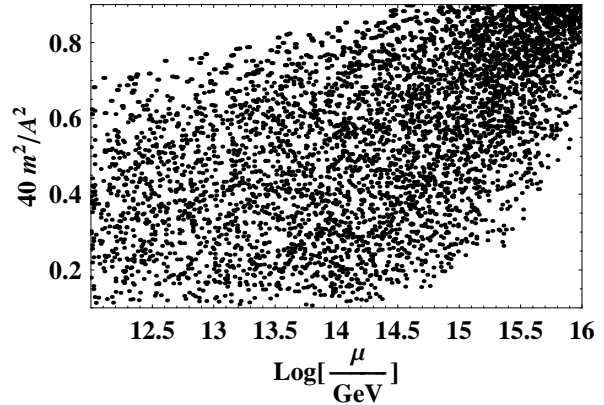}
\end{center}
\vskip -0.1in
\caption{The dots show the scale $\mu$ (horizontal axis) at which $A^2 = 40 m^2_\phi$, with the vertical axis representing the value of $40 m^2_\phi/A^2$ at $M_{\rm GUT}$, in the case of $udd$ flat direction. It is seen that for a large number of points the scale is within the interesting range $10^{14}-10^{15}$ GeV.}
\label{udd3}
\end{figure}

\section{MSSM inflation in minimal supergravity}

In this section we discuss how the allowed regions of the parameter space of MSSM inflation can be identified within the mSUGRA scenario by including all of the known cosmological and phenomenological constraints. In mSUGRA the soft masses of scalars and gauginos (denoted by $m_0$ and $m_{1/2}$ respectively), and the trilinear $A$-terms (denoted by $A_0$~\footnote{Note that this is not directly related to the non-renormalizable $A$-term in Eq.~(\ref{scpot}), which we consider in the previous sections.}) are set to be universal at an input energy scale, typically chosen as $M_{\rm GUT}$.
There are two more parameters in mSUGRA coming from the electroweak symmetry breaking sector: the ratio of the two Higgs VEVs, $\tan \beta$, and the sign of the Higgs mixing parameter. In this setup, we can use the RGEs to get the spectrum at any scale between the GUT scale and the weak scale.

Here we assume the lightest supersymmetric particle (LSP), which is stable due to $R$-parity conservation and therefore the dark matter candidate, to be a neutralino (denoted by ${\tilde \chi^0_1}$). The mSUGRA with a neutralino dark matter has been studied extensively in the literature. It is well known that only certain regions of the parameter space can satisfy the dark matter relic density range suggested by the WMAP data: (1) the bulk region with small $m_{1/2}$ and $m_0$, (2) the neutralino-stau coannihilation region, (3) the Higgs funnel regions, (4) the focus point (FP)/hyperbolic branch (HB) region, and (5) the neutralino-stop coannihilation region.

Neutralino dark matter is also subject to direct searches. There are many experiments devoted for the search of dark matter particles, including CDMS~\cite{CDMS}, DAMA~\cite{DAMA}, XENON~\cite{XENON}, EDELWEISS~\cite{EDELWEISS}, CRESST~\cite{cresst}, DEAP~\cite{deap}, CLEAN~\cite{clean}, LUX~\cite{lux}, EURECA~\cite{eureca}, and many more. CDMS has currently published their result, setting an upper limit of $3.8 \times 10^{-8}$ pb on the neutralino-proton spin-independent scattering cross section, $\sigma_{\tilde\chi^0_1-p}$, for a neutralino mass of 70~GeV~\cite{CDMSII}. Similar bounds have also been obtained by the XENON experiment~\cite{apriletalk}.

Other important constraints come from Br($b \to s \gamma$), Br$(B_s\rightarrow\mu^- \mu^+)$ (for large $\tan \beta$), the Higgs mass bound from LEP ($m_h > 114.4$~GeV~\cite{higgs1}), and the muon anomalous magnetic moment $g_\mu-2$~\footnote{Recent re-analysis by Davier {\it et al.}~\cite{Davier} of hadronic vacuum polarization contribution suggests a 3.2$\sigma$ 
deviation of $g_\mu - 2$ from the Standard Model, while a competing analysis by Teubner {\it et al.}~\cite{Teubner} suggests a 4$\sigma$
deviation. The $2 \sigma$ ranges of the deviation are $(9.5 - 41.5) \times 10^{-10}$~\cite{Davier} and $(15.8 - 47.4) \times 10^{-10}$~\cite{Teubner} respectively. We assume that this discrepancy comes from supersymmetry.}.

In a recent paper~\cite{darkcdms}, it was shown that we can satisfy all these constraints in parts of the neutralino-stau coannihilation region, where $\sigma_{\tilde\chi^0_1-p}$ is around $10^{-9} - 10^{-8}$~pb. The bulk region is generally already ruled out. The focus point region, while can satisfy all the other constraints, is not favored by the $g_\mu -2$. Similarly for the funnel region where, in addition, $\sigma_{\tilde\chi^0_1-p}$ is still orders of magnitude below the current limit. The neutralino-stop coannihilation requires a large value of $A_0$. At $A_0$ around 1~TeV this region is excluded by the  Br($b \to s \gamma$) constraint. At larger $A_0$ it becomes available, but no longer favored by the $g_\mu -2$~\cite{stopco} (except for a narrow region if we use the smaller lower limit of $g_\mu -2$~\cite{Davier}).

We now look at the allowed regions of the parameter space of MSSM inflation when all the bounds are included.

\subsection{Combined bounds from dark matter and inflation}

Bounds from the dark matter, and other bounds from phenomenology, are usually plotted in the $m_0-m_{1/2}$ plane. For any region in this plane one can find the corresponding region in the $m_\phi-\phi_0$ plane by using the relevant RGEs for the $udd$ and $LLe$ flat directions. 

\begin{figure}
\begin{center}
\includegraphics[width=.48\textwidth]{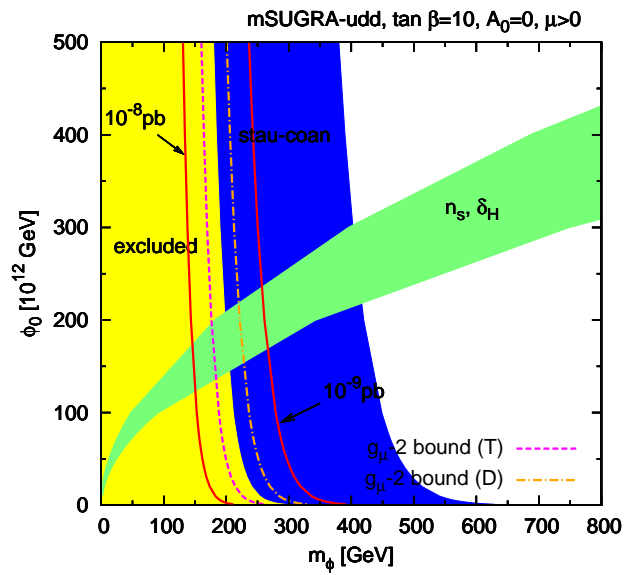}
\includegraphics[width=.48\textwidth]{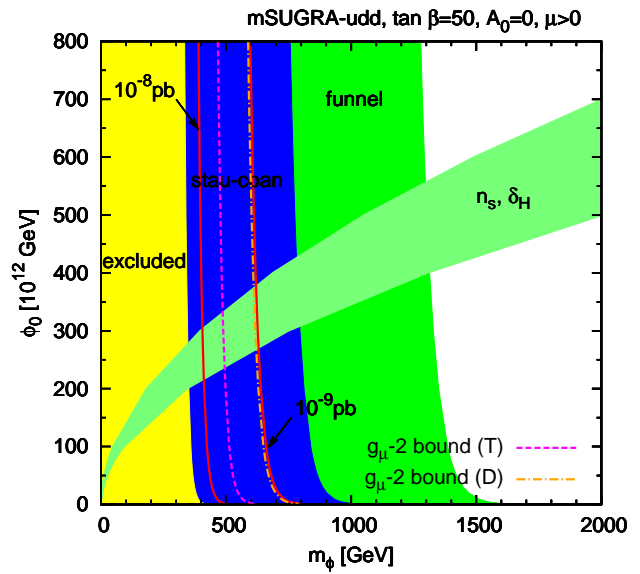}
\end{center}
\vskip -0.1in
\caption{The $m_\phi-\phi_0$ plane for the mSUGRA scenario, assuming $udd$ flat direction as the inflaton, with $A_0 = 0$ and ${\rm tan} \beta = 10~(50)$ in the upper (lower) panel. The $g_\mu - 2$ bounds, denoted by D and T, are from~\cite{Davier} and~\cite{Teubner} respectively. The contours and shadings are described in the text.}
\label{Fig:msugra_udd}
\end{figure}

In Fig.~\ref{Fig:msugra_udd}, we show the $m_\phi-\phi_0$ plane for the mSUGRA in the case of $udd$ flat direction, with $A_0 = 0$ and ${\rm tan} \beta = 10~(50)$ in the upper (lower) panel. The light green diagonal band shows the $2 \sigma$ allowed range for $n_s$ and $\delta_H$, the same as in Fig.~\ref{nsdH}. The blue shaded area corresponds to the neutralino-stau coannihilation region. The yellow shaded area is excluded by various phenomenological constraints (the strongest one coming from the LEP bound on the Higgs mass $m_h > 114.4$~GeV). The intersection between the light green (from inflation) and blue (from dark matter) bands represents the acceptable region of the $m_\phi-\phi_0$ plane. The $g_\mu - 2$ bounds can be used to further narrow down this region. The lower value in the $2 \sigma$ allowed range of $g_\mu-2$ is represented by the orange dot-dashed line and the pink dashed line, according to~\cite{Davier} and~\cite{Teubner} respectively, in the figure. The region to the left of these lines can provide a supersymmetric explanation of the $g_\mu-2$.

For ${\rm tan} \beta = 10$, we see that a small part of the coannihilation region is compatible with the $g_\mu - 2$ result in~\cite{Davier}, but the whole coannihilation region is disfavored if we use the result in~\cite{Teubner}. When all constraints are included, and using~\cite{Davier}, we find a rather narrow region that is allowed: $m_\phi \simeq 200-250$ GeV and $\phi_0 \sim (1.3-2.2) \times 10^{14}$~GeV. For this region, $\sigma_{\tilde\chi^0_1-p}$ is between $10^{-9}$~pb and $10^{-8}$~pb (shown by red contours).

For $\tan \beta = 50$, in addition to the neutralino-stau coannihilation region there is also the funnel region, which corresponds to the green shaded region~\footnote{The focus point/hyperbolic branch region in mSUGRA requires large $m_0$ and this leads to large values of $m_\phi$, which overlaps with the $m_\phi$ range of the funnel region and can extend to several TeV in some parts of the parameter space. It is not favored by the $g_\mu -2$ bound, and  we do not plot it here.}, which corresponds to the green shaded region. The strongest exclusion in this case comes from Br$(B_s\rightarrow\mu^- \mu^+)$ constraint. Note that only the coannihilation region is compatible with the $g_\mu-2$ bound. With all constraints taken into account, we find an allowed region: $m_\phi \simeq 350-600$~GeV and $\phi_0 \sim (2-3.5) \times 10^{14}$~GeV.
In this case $\sigma_{\tilde\chi^0_1-p}$ is between $10^{-9}$ pb and just above $10^{-8}$~pb.

\begin{figure}
\begin{center}
\includegraphics[width=.48\textwidth]{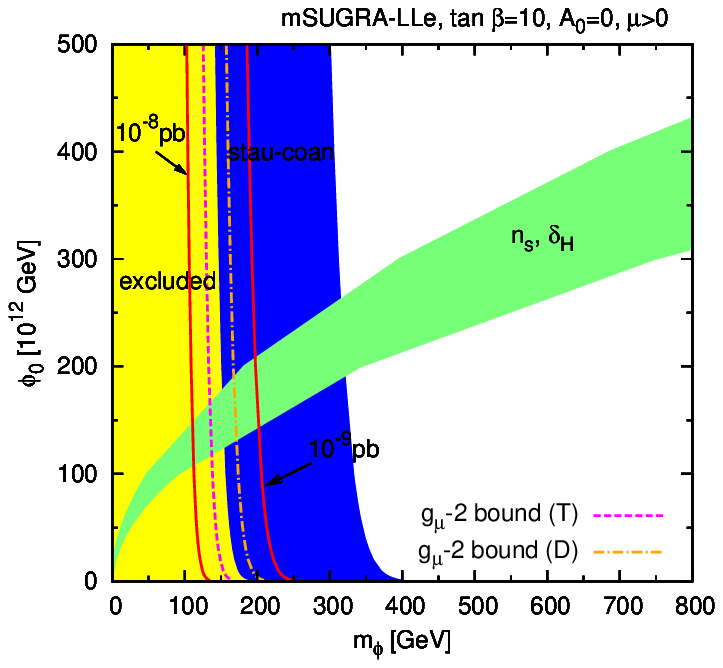}
\includegraphics[width=.48\textwidth]{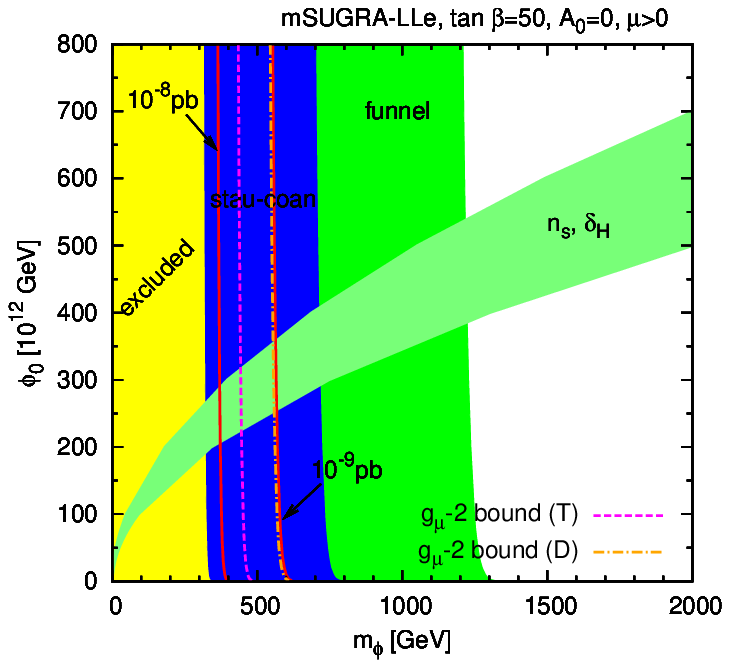}
\end{center}
\vskip -0.1in
\caption{The same as Fig.~\ref{Fig:msugra_udd} for $\tan \beta = 10(50)$ and $A_0 = 0$, assuming $LLe$ flat direction as the inflaton. }
\label{Fig:msugra_lle}
\end{figure}

Similar figures, Fig.~\ref{Fig:msugra_lle}, are obtained in the case of $LLe$ flat direction.
Again, when all constraints are included, we find a narrow region that is allowed. For $\tan \beta = 10$: $m_\phi \simeq 150-170$ GeV and $\phi_0 \sim (1.3-1.8) \times 10^{14}$~GeV. For this region, $\sigma_{\tilde\chi^0_1-p}$ is between $10^{-9}$~pb and $10^{-8}$~pb (shown by red contours). For $\tan\beta=50$ the allowed region is: $m_\phi \simeq 320-550$ GeV and $\phi_0 \sim (1.9-3.5) \times 10^{14}$~GeV. In general, the $LLe$ case yields smaller $m_\phi$ compared to the $udd$ for the same value of $\tan \beta$.

It is interesting to note the complementarity between the bounds from inflation and dark matter/phenomenology. The latter essentially lead to vertical bands and contours in the $\phi_0-m_\phi$ plane, which allows a significant restriction of inflation parameter space (the green band). This can be intuitively understood as follows. The vertical axis $\phi_0$, i.e. the inflection point VEV, plays a crucial role in inflation. Regarding dark matter and phenomenology, however, it is merely a scale at which the model parameters can be related to the weak scale observables, or to the input parameters at the GUT scale, by using the relevant RGEs.

We note that dark matter bounds alone are not strong enough to tightly constrain the $\phi_0-m_\phi$ plane. They allow for different regions in which the neutralino-proton elastic scattering cross section can be the same. One also needs to use bounds from other measurements to exclude some of these regions. As we saw, this can happen when the $g_\mu - 2$ bounds are taken into account. Another possibility is to use the low energy spectrum of the model, which we discuss next.

\subsection{Combined bounds from LHC and inflation}

SUSY particles will also be seen at colliders provided that they are not too heavy. The LHC is expected to probe squarks and gluino with  masses up to 3 TeV, which is a suitable mass range for MSSM inflation. Moreover, the LHC should be able to measure some of the masses to a very high accuracy. It has been shown~\cite{Bhaskar-LHC} that for the neutralino-stau coannihilation region in mSUGRA the LHC measurements can be used to determine the SUSY parameters very accurately.

\begin{figure}
\begin{center}
\vskip 0.2in
\includegraphics[width=.48\textwidth]{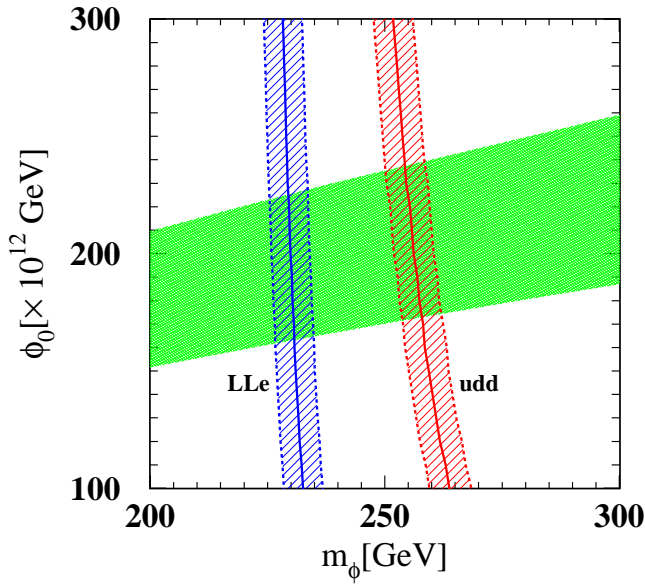}
\end{center}
\vskip -0.15in
\caption{The allowed regions in the $m_\phi-\phi_0$ plane as a result of mass measurement at the LHC, for the mSUGRA scenario with $\tan \beta = 40$ and $A_0 = 0$. Here, $m_0 = 210 \pm 4$~GeV and $m_{1/2} = 350 \pm 4$~GeV, with uncertainties due to experimental uncertainty in the masses.}
\label{Fig:LHC}
\end{figure}

In Fig.~\ref{Fig:LHC} we show the $m_\phi-\phi_0$ plane for $\tan \beta = 40$, $A_0 = 0$, $m_0 = 210 \pm 4$~GeV and $m_{1/2} = 350 \pm 4$~GeV, which can be obtained from mass measurements at the LHC for $10$ fb $^{-1}$ of luminosity~\cite{Bhaskar-LHC}. The uncertainties in $m_0$ and $m_{1/2}$ correspond to small experimental uncertainty in the measured masses, which are translated into narrow bands in the $m_\phi-\phi_0$ plane by using the relevant RGEs. The allowed region is the intersection of these bands (which come from mass measurement at the LHC) and the green band (coming from the CMB constraints on density perturbations).
We see that the allowed range for $m_\phi$ depends on which of the $LLe$ or $udd$ flat directions plays the role of the inflaton.
In the case of $udd$ flat direction $m_\phi \simeq 250-260$~GeV is obtained, while in the case of $LLe$ flat direction we find $m_\phi \simeq 225 - 235$~GeV. In both cases we have $\phi_0 \sim (1.6-2.2) \times 10^{14}$~GeV.

It is seen that the bands corresponding to the bounds from mass measurements are almost vertical with a slight tilt to the left. This is due to the fact that the soft masses of squarks and sleptons decrease at higher scale (larger $\phi_0$) due to the dominance of gaugino loops. The effect is less prominent for sleptons (hence $LLe$ flat direction) because of the absence of gluino loops, which is why we have a practically vertical band in this case. Again the mass measurements do not restrict $\phi_0$ as it does not play any role in phenomenology. Obtaining tighter constraints on the allowed range of $\phi_0$ requires a more precise measurement of the scalar spectral index $n_s$ in the CMB experiments, like PLANCK.

More data in the future will allow us to shrink the bands and, eventually, pin down a tiny region in the $m_\phi-\phi_0$ plane. Once we know the inflection point VEV and the inflaton mass, then we can use Eq.~(\ref{infvev}) to find the dimensionless coupling $\lambda$. Also, the non-renormalizable $A$-term will be found from Eqs.~(\ref{dev},\ref{pot},\ref{tilt},\ref{Delta},\ref{NCOBE}). The values of $m_\phi,~\lambda,~A$ thus obtained are all at the scale $\phi_0$, and can be related to the input parameters at $M_{\rm GUT}$ by using the relevant RGEs. We emphasize that $\lambda$ and $A$ have no bearing for phenomenology, but are crucial in finding an inflection point in the potential at the right scale.

\section{Conclusion}

MSSM inflation represents a realistic embedding of inflation in high energy physics where the inflaton has a natural place in a well-motivated and testable model of particle physics instead of being added as an extra field. In this paper we discussed some aspects of MSSM inflation mainly focusing on two main issues.

The robustness of MSSM inflation with regard to its predictions for density perturbations, which is due to the fact that inflation occurs near a point of inflection, is a remarkable feature. However, generating acceptable perturbations requires that a very precise relation between the soft SUSY breaking parameters be satisfied up to several orders in perturbation theory.

We suggested a different perspective on this issue.
The necessary relationship can be satisfied at scales, which are phenomenologically interesting, without any fine tuning between the boundary values of the soft SUSY breaking parameters at an input scale (like the GUT scale). For given boundary values, after using the relevant RGEs, we can find the scale at which the relation is satisfied (Figs.~2,~3,~4). Requiring that this scale matches a point of inflection in the potential, determines a dimensionless coupling that represents a non-renormalizable interaction term in the superpotential to very high accuracy. This coupling has no bearing for phenomenology and its only role is to give rise to a point of inflection that is suitable for inflation.

Another important feature of MSSM inflation, due to the fact that the inflaton is related to squarks and sleptons, is that it can be tested outside cosmology. Once a specific framework is supposed, we can obtain predictions of the inflationary model for phenomenology. Then the CMB measurements can be combined with various phenomenological bounds to identify the allowed parameter space of the model.

We presented a detailed study of the parameter space of MSSM inflation in the case of mSUGRA scenario. We demonstrated the complementarity between the bounds from CMB measurements and those from phenomenology (Figs.~5,~6,~7) in the $m_\phi-\phi_0$ plane ($\phi_0$ and $m_\phi$ denoting the point of inflection VEV and the inflaton mass calculated at that scale respectively). The limits from SUSY dark matter and future mass measurement of SUSY particles at the LHC, as well as muon anomalous magnetic moment and rare decays, significantly restrict the allowed range of $m_\phi$. On the other hand, more precise determination of the scalar spectral index from CMB experiments will further narrow down the allowed range of $\phi_0$.

More data from a whole array of experiments (PLANCK, LHC, dark matter direct detection experiments, etc) will lead to tighter constraints on the allowed region of the parameter space. Eventually, this collaboration between cosmology and particle physics can be used to determine the underlying parameters of the MSSM inflation.

\section{Acknowledgements}

The work of B.D. is supported in part by the DOE grant DE-FG02-95ER40917. The work of Y.S. is supported in part by the DOE grant DE-FG02-04ER41308.

\end{document}